\documentclass[a4paper]{jacow}
\usepackage{siunitx}
\usepackage[T1]{fontenc}
\usepackage{graphicx}
\usepackage{stfloats}
\usepackage{hyperref}
\newcommand{\doi}[1]{\href{https://doi.org/#1}{https://doi.org/#1}}
\graphicspath{{figures/png/}}

\newcommand{\safeincludegraphics}[2][]{%
\IfFileExists{#2}{%
\includegraphics[#1]{#2}%
}{%
\IfFileExists{figures/pdf/#2}{%
\includegraphics[#1]{figures/pdf/#2}%
}{%
\fbox{\parbox[c][0.16\textheight][c]{0.94\linewidth}{\centering
Missing figure: \texttt{#2}\
Place the PDF in the current directory or in \texttt{figures/pdf/}.}}%
}%
}%
}

\title{Development and Testing of Tools for ORM Measurements and Quadrupole
Centering}

\author{A.~Pathak\thanks{This work was produced by FermiForward Discovery Group, LLC
under Contract No.\ 89243024CSC000002 with the U.S. Department of Energy, Office of
Science, Office of High Energy Physics.}\thanks{abhishek@fnal.gov}, E.~Chen, A.~Shemyakin \\
Fermi National Accelerator Laboratory, Batavia, IL, USA}

\begin{document}
\maketitle

\begin{abstract}
In preparation for the commissioning of Linac2 (formerly referred to as the {PIP-II}
linac) at Fermilab, we are developing high-level applications that automate Orbit Response
Matrix (ORM) measurement and beam-based localization of quadrupole magnetic centers.
Both procedures are designed to run in a low-amplitude mode compatible with otherwise
regular operation. The ORM is measured by exciting multiple dipole correctors
simultaneously at low amplitude, each at a distinct frequency, and applying a Discrete
Fourier Transform to the recorded BPM signals to extract per-element transfer functions
with quantified noise. One application of the measured ORM is to predict the corrector
coefficients that form local orbit perturbations (bumps). To center the trajectory in a
quadrupole, such a bump is created around it, the quadrupole current is oscillated, and
the responses of the downstream BPMs are recorded and analyzed. Repeating this at
several beam positions inside the quadrupole -- each set by the bump -- the dependence
of the BPM signal on beam position provides information about the quadrupole center. The
applications are being tested at the Fermilab 400\,MeV Linac during regular production runs,
and the lessons learned are fed back into their improvement.
\end{abstract}

\section{INTRODUCTION}
Commissioning of Linac2~\cite{pip2tdr,napoly_linac2026} will require rapid and repeatable beam-based optics diagnostics, including orbit response matrix (ORM)~\cite{safranek_loco} measurements, orbit correction, and trajectory centering through quadrupoles. During commissioning, these measurements can be performed during dedicated machine time. For the present development and testing at the Fermilab 400\,MeV Linac, however, we seek to perform the measurements parasitically during routine production operation, avoiding disruption to beam delivery. We therefore apply low-amplitude multi-frequency ac excitation~\cite{rehm_multifreq,marti_bba,shemyakin_napac2019,shemyakin_napac2025}, which allows multiple correctors to be excited simultaneously while the resulting beam response is extracted from the BPM signals. Two high-level applications were tested in this mode. FORMA drives all selected dipole correctors simultaneously, each at its own frequency, and demodulates the BPM signals to measure the ORM. AutoCenter uses the measured ORM to form local trajectory bumps and locate quadrupole centers from the downstream response. We also study orbit correction under ORM mismatch by comparing singular-value-decomposition (SVD) correction based on the measured ORM with Bayesian optimization (BO), which does not use the supplied global ORM.

\section{ORM MEASUREMENT METHOD}
\begin{figure*}[tb]
\centering
\includegraphics[width=\textwidth]{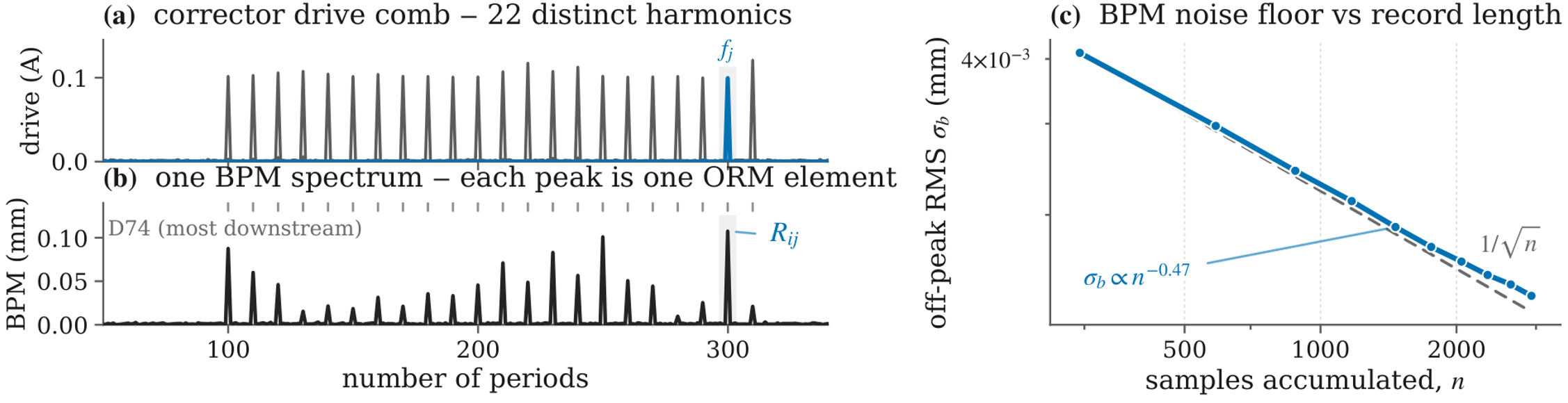}
\caption{FORMA ORM measurement, June 2, 2026 horizontal scan. Panels (a,b) use
the number of periods over the record, $q=fT_{\rm rec}\simeq k$, rather than hertz.
(a) Simultaneous corrector drive, one harmonic per corrector, with a single
corrector highlighted as an example; (b) amplitude spectrum of a single BPM
(D74, the most downstream), with each peak providing $\tilde b$ for an ORM
element ($R_{ij}$ obtained from $\tilde b/\tilde c$), the highlighted peak being
the response to the corrector marked in (a); (c) BPM noise floor
$\sigma_b\propto n^{-0.47}$, close to the ideal $1/\sqrt{n}$, which determines the
statistical uncertainty in Eq.~\eqref{eq:err}.
}
\label{fig:comb}
\end{figure*}

Each of the 22 horizontal (or 22 vertical) corrector channels is driven simultaneously with a sinusoidal excitation added to its operating setpoint:
\begin{equation}
x_j[n] = x_j^0 + A_j \sin\!\left(2\pi P_j \frac{n \bmod S}{S}\right),
\label{eq:drive}
\end{equation}
where $n$ is the sequence index and each corrector $j$ carries a distinct integer harmonic number $P_j=10,11,\dots,31$. The $S$-sample pattern is repeated ten times, so that over the full record of $N=10S$ samples each corrector completes $10P_j$ cycles and occupies the distinct DFT bin $k_j=10P_j$. Because the number of samples in one period, $S=293$, is prime, every $P_j$ is coprime with $S$; each drive waveform therefore has a period of exactly $S$ samples and repeats exactly ten times over the record, as assumed above. All correctors are driven simultaneously, and all 70 BPM channels are recorded synchronously on a common 5\,Hz beam pulse, with $A_j=0.1$\,A and $N=2930$. Although fewer samples were sufficient to reach the required noise level, the longer acquisition was retained deliberately to demonstrate the expected statistical reduction of the noise floor with record length. Because the beam-pulse-locked scan contains timing gaps, spectra are shown versus the number of periods completed over the record rather than frequency. For mean timestamp spacing $T$, elapsed duration $T_{\rm rec}\simeq NT$, and $f_k=k/(NT)$, the plotted coordinate is $q_k=f_kT_{\rm rec}=kT_{\rm rec}/(NT)\simeq k$. Thus the peaks occur at $q_j\simeq k_j=10P_j=100,110,\dots,310$.
At drive bin $k_j$, let $\tilde b_i$ and $\tilde c_j$ be the complex one-sided DFT coefficients of the mean-subtracted BPM $i$ and corrector-readback $j$ signals. The signed in-phase transfer coefficient is
$R_{ij}=\operatorname{Re}(\tilde b_i/\tilde c_j)$ in \si{mm/A}; division by $\tilde c_j$ normalizes the BPM response to the measured corrector amplitude and phase. Its statistical uncertainty is propagated from the off-peak spectral RMS of both signals:
\begin{equation}
\Delta R_{ij} = \sqrt{\frac{\sigma_b^2}{2A_c^2}
+ \frac{A_b^2\sigma_c^2}{2A_c^4}},
\label{eq:err}
\end{equation}
where $A_b=|\tilde b_i|$ and $A_c=|\tilde c_j|$ are the measured amplitudes, and $\sigma_b$ and $\sigma_c$ are the corresponding off-peak noise floors. Equation~\eqref{eq:err} assumes independent, isotropic complex noise, with each quadrature carrying $\sigma^2/2$; therefore, $\Delta R$ represents statistical uncertainty only. Figure~\ref{fig:comb} shows the simultaneous drive, the resulting BPM spectral comb, and the measured $\sigma_b\propto n^{-0.47}$ decrease of the noise floor with accumulated samples, close to the ideal $1/\sqrt{n}$, which motivates the selected record length.

\begin{figure*}[tb]
\centering
\includegraphics[width=\textwidth]{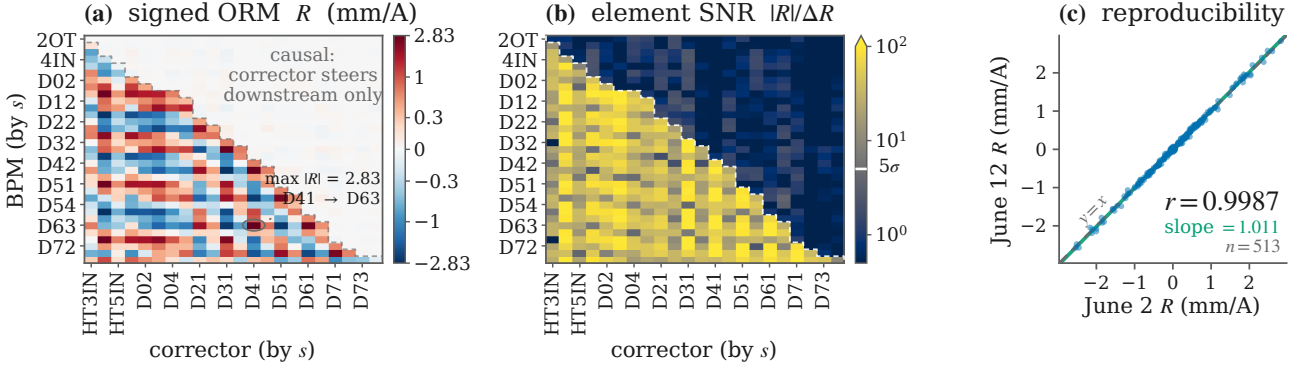}
\caption{Measured horizontal ORM and validation, with rows and columns ordered
by longitudinal position $s$. (a) Signed response $R$ (asinh scale,
$\max|R|=2.83$\,\si{mm/A}); (b) per-element SNR $|R|/\Delta R$, showing the
causal lower-triangular structure and downstream terms exceeding $5\sigma$;
(c) the 513 elements common to the two scans, showing agreement with the
June 12 repeat ($r=0.9987$, slope $=1.011$).}
\label{fig:orm}
\end{figure*}

\section{ORM RESULTS AND VALIDATION}
The measured horizontal ORM (Fig.~\ref{fig:orm}) is $33\times22$
(BPM$\times$corrector; 33 of 35 horizontal BPM channels enter the matrix), with
$\max|R|=2.83$\,\si{mm/A}. Ordered by longitudinal position, it exhibits the causal
structure expected of a transport line: a corrector produces a response only at
downstream BPMs. The RMS response in the downstream block is 62 times that in the
upstream block. The per-element SNR from Eq.~\eqref{eq:err} separates into two distinct groups:
96\% of the terms in the causal block exceed $5\sigma$, while the acausal terms
remain at the noise floor.

No design matrix was available, so validation rests on reproducibility and causality.
A repeat measurement ten days later reproduced the ORM closely. Over the
$27\times19=513$ elements common to both scans, the matrices agree with a Pearson
correlation coefficient of $r=0.9987$ and a slope of 1.011
(Fig.~\ref{fig:orm}(c)).

The drive amplitude was deliberately kept small to minimize perturbation of routine
operation. The $0.1$\,A drive produced a median per-BPM orbit excursion of 0.61\,mm --- the
combined effect of all 22 correctors, not a single-corrector response --- within
operational tolerances.

\begin{figure}[tb]
\centering
\includegraphics[width=\columnwidth]{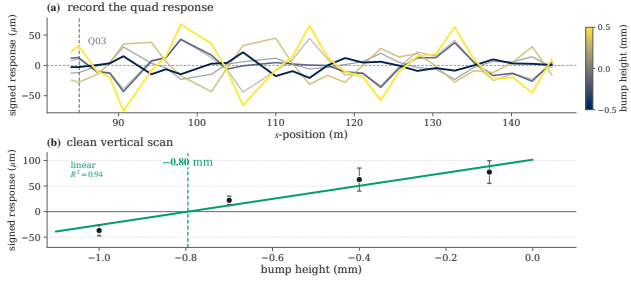}
\caption{Beam-based quadrupole centering (AutoCenter) on Q03, a representative
quadrupole of the Fermilab 400\,MeV Linac. (a) Signed BPM
response to quadrupole modulation along the beamline, colored by the imposed bump
height; (b) signed response against bump height for the clean vertical scan, with
a linear fit locating the center at its zero crossing. Error bars denote bootstrap
$1\sigma$ over the BPMs, excluding correlated jitter and calibration
uncertainties.}
\label{fig:center}
\end{figure}

\begin{figure}[tb]
\centering
\includegraphics[width=\columnwidth]{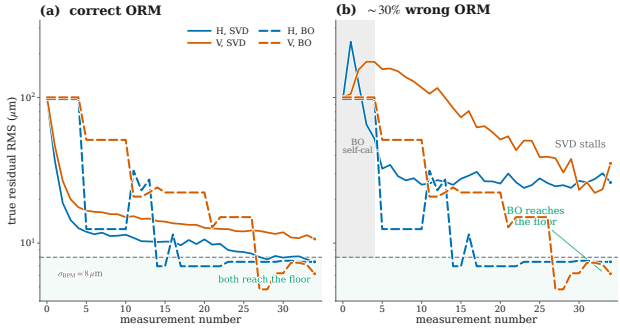}
\caption{Orbit correction with (a) the correct ORM and (b) an ORM about 30\% wrong.
Solid curves are Tikhonov-regularized SVD, dashed are Bayesian optimization; blue is
horizontal, orange vertical. $\sigma_{\rm BPM}=8$\,\si{\micro\metre} marks the
per-BPM noise floor.}
\label{fig:orbit}
\end{figure}

\section{BEAM-BASED QUADRUPOLE CENTERING}
A quadrupole produces a downstream orbit change when the beam is offset from its magnetic
center; modulating its strength therefore produces a response proportional to the beam
offset that changes sign as the beam crosses the center. From the measured ORM, AutoCenter
uses regularized least squares to form a local trajectory bump from neighboring
correctors. The method is demonstrated on Q03, a representative quadrupole of the
Fermilab 400\,MeV Linac, and applies to any quadrupole with a downstream BPM and
enough neighboring correctors to close a local bump. The scan coordinate $h$ is the requested displacement at the BPM located
0.1\,m from Q03 and serves as a proxy for the beam offset through the quadrupole. At
each value of $h$, the quadrupole current is sinusoidally modulated with a 1\,A amplitude,
and a DFT extracts the signed, in-phase component of the response at each downstream BPM.
The responses vary approximately linearly with $h$, so a linear fit to the data locates
the beam center at its zero crossing (Fig.~\ref{fig:center}).

Because the measured ORM is imperfect, the resulting bump does not close exactly and
leaves a small residual trajectory. A closed-loop leak-correction step therefore
re-flattens the non-target BPMs while preserving the requested displacement at the target
BPM. For Q03, the bump used five correctors, D01--D04 and D11. The leak-correction loop used a
gain of 0.4, a 0.1\,mm convergence threshold, and no more than five iterations. The
quadrupole modulation was applied for two periods in the 100-point production scan and
ten periods in the 293-point clean scan.

A vertical scan of Q03 produced a well-resolved center
(Fig.~\ref{fig:center}(b)). A linear fit to the signed response against bump height
crosses zero at $-0.80$\,mm, with $R^2=0.94$. The error bars are the bootstrap
$1\sigma$ over the BPMs and do not include correlated jitter or calibration
uncertainties. Because no independent survey or magnetic-axis
measurement of Q03 was available, the result represents a relative center with respect
to the baseline orbit rather than an absolute magnetic-center measurement.

\section{ORBIT-CORRECTION ROBUSTNESS MEASUREMENTS}
In closed-loop measurements on the Linac, correction using the measured ORM with a
Tikhonov-regularized SVD controller (gain 0.7) was compared with Bayesian
optimization (BO)~\cite{duris_bo,gao_bo_traj}. BO does not use the supplied global ORM;
its first four measurements build a local response basis from a baseline and one jog per
variable. In each plane, both methods use the same three bump correctors and an equal
34-measurement online budget to correct a 0.5\,mm-peak bump at D03
($\approx0.10$\,mm orbit RMS) with $8$\,\si{\micro\metre} rms BPM noise
(Fig.~\ref{fig:orbit}(a)). SVD uses
$R^{\mathsf T}(RR^{\mathsf T}+\alpha I)^{-1}$ with
$\alpha=0.01\,(\si{mm/A})^2$; BO uses a Mat\'ern-$5/2$ Gaussian process~\cite{rasmussen_gpml} with
expected-improvement acquisition. Of BO's 34 measurements, four build the local basis,
six are quasi-random, and 24 use expected improvement. The SVD curve shows the applied
iterate, while the BO curve shows the best measured setting.

Residuals are medians over the final eight measurements of the true residual, the
noise-free orbit RMS over all BPMs computed with the true response matrix. With the
nominal matrix, SVD reaches $8.0$\,\si{\micro\metre} (H) and
$11.6$\,\si{\micro\metre} (V), while BO reaches $7.5$\,\si{\micro\metre} (H) and
$6.2$\,\si{\micro\metre} (V), all within a factor 1.5 of the
$8$\,\si{\micro\metre} per-BPM noise level, which sets the scale of what either
method can resolve. To assess sensitivity
to ORM mismatch, the machine was left unchanged while the SVD controller was supplied
with a perturbed matrix,
$R^{\rm ctrl}_{ij}=R^{\rm true}_{ij}(1+\epsilon_{ij})(1+\delta_j)$, where
$\epsilon\sim\mathcal N(0,0.18^2)$ and $\delta\sim\mathcal N(0,0.22^2)$.
The resulting relative matrix mismatch,
$\eta_R=\lVert R^{\rm ctrl}-R^{\rm true}\rVert_F/
\lVert R^{\rm true}\rVert_F$, was 0.22 in H and 0.33 in V for the controller columns
and approximately 0.30 over the full matrix. Under this mismatch, SVD stalled at
$26$\,\si{\micro\metre} in both planes. The perturbation enters only the SVD controller
matrix, so the BO curve in (b) is a repeat run serving as a reference, not a mismatch
test; BO reaches the noise level without a global ORM. The comparison
uses the same initialization and noise realization for both cases; one perturbation
realization is considered.

\section{CONCLUSION}
Both FORMA and AutoCenter ran on the Fermilab 400\,MeV Linac, including during production operation.
The ORM was reproducible within system drift, and a vertical scan produced a
well-resolved relative quadrupole center. The preliminary production scan established practical requirements for DFT
integration and leak-loop convergence.
In the single seeded perturbation realization studied here, SVD stalled under
controller-matrix mismatch, while BO, which never uses the supplied matrix, again
reached the noise level. The applications are now being
adapted for Linac2 commissioning.



\end{document}